\documentclass{edm_article}

\usepackage{listings}
\usepackage[table,xcdraw]{xcolor}
\usepackage{amsmath}
\usepackage{hyperref}
\usepackage[numbers]{natbib}
\begin{document}

\title{pyBKT: An Accessible Python Library of Bayesian Knowledge Tracing Models}

\numberofauthors{3} 
\author{
\alignauthor
 Anirudhan Badrinath\\
    \affaddr{University of California, Berkeley, CA, USA}\\
        \email{abadrinath@berkeley.edu}
  \alignauthor Frederic Wang\\
        \affaddr{University of California, Berkeley, CA, USA}\\
        \email{fredwang@berkeley.edu}
  \alignauthor
 Zachary Pardos\\
        \affaddr{University of California, Berkeley, CA, USA}\\
        \email{pardos@berkeley.edu}


}

\maketitle


\begin{abstract}
Bayesian Knowledge Tracing, a model used for cognitive mastery estimation, has been a hallmark of adaptive learning research and an integral component of deployed intelligent tutoring systems (ITS). In this paper, we provide a brief history of knowledge tracing model research and introduce pyBKT, an accessible and computationally efficient library of model extensions from the literature. The library provides data generation, fitting, prediction, and cross-validation routines, as well as a simple to use data helper interface to ingest typical tutor log dataset formats. We evaluate the runtime with various dataset sizes and compare to past implementations. Additionally, we conduct sanity checks of the model using experiments with simulated data to evaluate the accuracy of its EM parameter learning and use real-world data to validate its predictions, comparing pyBKT's supported model variants with results from the papers in which they were originally introduced. The library is open source and open license for the purpose of making knowledge tracing more accessible to communities of research and practice and to facilitate progress in the field through easier replication of past approaches.
\end{abstract}

%


\section{Introduction}
Knowledge Tracing \cite{corbett1994knowledge} has been a well researched approach to estimating students' cognitive mastery in the context of computer tutoring systems \cite{pelanek2017bayesian}. Tutoring systems take a problem-solving, or active approach to learning \cite{anzai1979theory,aleven2002effective} that often resembles the personalized mastery learning approach researched by \citet{bloom19842}. The model was not originally described using a particular statistical framework; however, the mathematical expressions in the original work are consistent with Bayes Theorem \cite{reye2004student}, and the canonical model was subsequently coined Bayesian Knowledge Tracing (BKT). 

In spite of its growing popularity in the research community, accessible and easy to use implementations of the model and its many variants from the literature have remained elusive. In this paper, we introduce pyBKT\footnote{\url{https://github.com/CAHLR/pyBKT}}, a modernized Python-based library, making BKT models and respective adaptive learning research more accessible to the community. The library's interface and underlying data representations are expressive enough to replicate past BKT variants and allow for new models to be proposed. The library is designed with data helpers and model definition functions, allowing for convenient replication and comparison to BKT model variants and subsequently better scientific progression and evaluation of new state-of-the-art knowledge tracing approaches.

The Bayesian Knowledge Tracing model can be described as a Hidden Markov Model (HMM) with observable nodes representing students’ known binary problem response sequences $obs_t$ and hidden nodes representing students’ latent knowledge state at a particular time step $t$. Using expectation maximization, pyBKT fits the learn (transmission parameter), and guess, and slip (emission) parameters from historical response logs, with the parameters defined below.
$$
\text{prior} = P(L_0)
$$
$$
\text{learn} = P(T) = P(L_{t+1} = 1 | L_t = 0)
$$
$$
\text{guess} = P(G) = P(obs_t = 1 | L_t = 0)
$$
$$
\text{slip} = P(S) = P(obs_t = 0 | L_t = 1)
$$
Note that while $P(L_0)$ denotes the prior parameter, we also define $P(L_t)$ as the probability that the student has mastered the skill at time step $t$. Bayesian Knowledge Tracing updates $P(L_{t})$ given an observed correct or incorrect response to calculate the posterior with:
$$P(L_t | obs_t = 1) = \frac{P(L_t)(1-P(S))}{P(L_t)(1-P(S))+(1-P(L_t))P(G)}$$
$$P(L_t | obs_t = 0) = \frac{P(L_t)P(S)}{P(L_t)P(S)+(1-P(L_t))(1-P(G))}$$
The updated prior for the following time step, which incorporates the probability of learning from immediate feedback and any other instructional support, is defined by:
$$P(L_{t+1}) = P(L_t|obs_t) + (1-P(L_t|obs_t))P(T)$$
The standard BKT model assumes no forgetting: $$P(F) = P(L_{t+1} = 0 | L_t = 1) = 0$$

\section{Related Work}


Since the introduction of the standard BKT model by \citet{corbett1994knowledge}, many variants of BKT have been proposed which multiplex, or condition the four parameters of prior, learn, guess, and slip on different factors such as question type or student. KT-IDEM (Item Difficulty Effect) \cite{pardos2011kt} captured item performance variance within skill by allowing different guess and slip values to be fit per item. \citet{d2008more} proposed a hybrid model using regression to help determine if a student's response was a guess or a slip based on context. 

Other modifications focused on conditioning the learn rate and prior knowledge parameter \cite{pardos2010modeling}. \citet{yudelson2013individualized} explored student-level parameter individualization, finding that the learn rate provided better predictive performance than prior when individualized to each student. Learn rate has also been conditioned based on the item \cite{pardos2009detecting}, educational resource (e.g., a video in an online course) \cite{pardos2013adapting}, or type of instructional intervention seen by the student at each opportunity \cite{lin2016intervention} or order in which items or resources were seen \cite{pardos2009determining}. The assumption of no forgetting was relaxed in \citet{qiu2011does}, finding that response correctness decreased based on time elapsed since the last response. This decrease was better modeled in BKT with conditional forget rates than with increased slip rates. \citet{gonzalez2014general} created a hybrid BKT model that factored in a variety of features, including response time. An overview of other variations of BKT and logistic approaches to learner response prediction can be found in \citet{pelanek2017bayesian}.


Neural network approaches to knowledge tracing have gained in momentum since the introduction of Deep Knowledge Tracing (DKT; \cite{piech2015deep}). Some have explored the reasons for DKT's apparent accuracy improvement as compared to BKT and have attributed its success to its high dimensional hidden space and ability to observe interleaved skills in a single model \cite{montero2018does}. However, most papers using neural approaches compare only to the standard BKT model proposed in the 1990s and not to the more modern variants. A noteworthy result of caution was reported in \citet{khajah2016deep}, in which it was found that simply enabling the forget parameter of standard BKT led to performance on par with DKT on several datasets.

There has been a brief history of BKT implementation frameworks. Several BKT variants have used Kevin Murphy's Bayes Net Toolbox (BNT) for MATLAB \cite{murphy2001bayes}, with a subsequent wrapper for that toolbox releases catering to knowledge tracing \cite{chang2006bayes}. \citet{yudelson2013individualized} produced a C++ implementation\footnote{https://github.com/myudelson/hmm-scalable} of BKT with a command line interface that included support for individualized parameters. Finally, \citet{xu2015xbkt} created a C++ implementation of BKT with a MATLAB interface and support for parallelization and conditioning of all parameters based on both problems and passive resources (e.g., a learning rate for a video).

\section{pyBKT Library Details}
The pyBKT library builds off of xBKT developed by \citet{xu2015xbkt} and is released under an MIT license. The library is compatible with all platforms (Linux, Windows, Mac OS), primarily utilizing NumPy \cite{walt2011numpy} for computation. It is available on the Python PyPi repository, with installation accomplished through a pip one-liner: \verb+pip install pyBKT+. To increase performance, we additionally supply routines which utilize C++ libraries and Eigen/LAPACK, which is an optimized linear algebra package with OpenMP support. This accelerated version requires a C++ compiler and is currently only tested on Linux and macOS.

We created a Scikit-learn style \cite{pedregosa2011scikit} Model class abstraction and accompanying data helpers that further facilitate the accessibility and expressive power of pyBKT. With one-line fit, predict, evaluate, parameter initialization, and cross-validate methods, pyBKT offers ease of use in ingesting response data and applying BKT models and supported variants. We explore the interface to these methods in the next subsection. We then detail the internal data structures, computations, and motivations behind the development of the two implementations of pyBKT, in pure Python and the accelerated C++/Python, along with runtime evaluation. 

\subsection{Interface}

pyBKT's interface is modeled after Scikit-learn's accessible frontend interface for machine learning models \cite{pedregosa2011scikit}. The ease-of-use in the pyBKT Model class abstraction allows for increasingly expressive BKT code without the usability sacrifices of past BKT libraries. We aim for the library to be easy to learn for beginners while still useful for experienced users conducting knowledge tracing research. Further, it provides a gateway into exploring multiple model extensions from the literature, which have been shown to be capable competitors to DKT \cite{khajah2016deep} and able to address inequities in unmodeled differences in learning and prior ability between students \cite{doroudi2019fairer}. Supported BKT extensions include: KT-IDEM \cite{pardos2011kt}, KT-PPS (Prior Per Student) \cite{pardos2010modeling}, BKT+Forget \cite{khajah2016deep}, Item Order Effect \cite{pardos2009determining} and Item Learning Effect \cite{pardos2009detecting,pardos2013adapting}. These model extensions are referred to as \verb+multigs+, \verb+multiprior+, \verb+forgets+, \verb+multipair+, and \verb+multilearn+, respectively, in the model interface.

The Model class abstraction supports creating, fitting, predicting, cross-validating and evaluating BKT models using any combination of supported extensions. Additional features include specifying model parameter initialization before fitting, custom cross-validation fold assignment, and multiple accuracy and error metrics - including support for generic user-defined or Scikit-learn imported metrics. Common dataset formats are made easier to ingest through automatic detection of familiar column headers seen in Cognitive Tutor \cite{koedinger1997intelligent} and ASSISTments datasets \cite{heffernan2014assistments}. Defaults for all customizable parameters such as random seed, parallelization, model variants, and evaluation metric(s) are provided when they are not specified.

We demonstrate a few of the library's basic capabilities in parameter initialization, fitting, and parameter output in the below code snippet using the learned parameters of the "Polynomial Factors" skill from the 2009-2010 ASSISTments dataset\footnote{\url{https://sites.google.com/site/assistmentsdata/home/assistment-2009-2010-data}}. Note that all parameters of all skills found in the dataset are fit unless otherwise specified.

\begin{lstlisting}[language=Python, breaklines=true, showstringspaces=false, columns=flexible, tabsize = 12]
>>> from pyBKT.models import Model
>>> model = Model()
>>> model.fit(data_path = 'assistments.csv')
>>> model.params().loc[('Polynomial Factors')]
param   class     value
prior   default 0.17452
learns  default 0.13378
guesses default 0.25502
# ...
\end{lstlisting}

The internal data helper functionality converts any response-per-row comma or tab separated file into the internal pyBKT data format. It is designed to convert input columns using default column mappings for skill name, student identifiers, correctness, etc. automatically for Cognitive Tutor/ASSISTments data and configurable with one line for any other dataset. It provides increased flexibility to the user, allowing for consistency across fit and predict/evaluate phases.

The included evaluation metrics are root-mean squared error (RMSE), accuracy with a threshold of 0.5, and area under the ROC curve (AUC). Custom metrics in the format of two-parameter Python functions are supported for evaluation and cross-validation, such as the regression or classification metrics from \verb+sklearn.metrics+ or \verb+keras.metrics+.

The cross-validate function provides a one-line interface for fitting and evaluating any combination of model variants with one or more error metrics. For the following example, we specify a particular column to use for the multilearn (answer type) along with a multigs model trained on the default template ID for ASSISTments. A skill or combination of skills can be specified along with the seed and number of folds (optional). 

\begin{lstlisting}[language=Python, breaklines=true, showstringspaces=false, columns=flexible, tabsize = 12]
>>> model = Model(seed = 42, parallel = True)
>>> model.crossvalidate(data_path = 'assistments.csv', skills = ['Circle Graph', 'Box and Whisker'], multigs = True, multilearn = 'answer_type', metric = ['auc', sklearn.metrics.mean_absolute_error])
skill            mean_absolute_error     auc                                       
Circle Graph                 0.41565 0.72782
Box and Whisker              0.33906 0.67991
\end{lstlisting}

\subsection{Internal Implementation Details}

The pure Python and C++/Python implementations of pyBKT both make use of optimized programmatic methods, efficient internal data and model representation, multithreaded model fitting, and optimized linear algebra libraries. The model fitting consists of a typical Expectation Maximization (EM) function for a Hidden Markov Model performing forward and backward passes over the sequential data to continuously update BKT parameters. We implement these passes using a parallelized iterative dynamic programming approach on the input data. 

\subsubsection{Model Representation}
In the context of model variants with multiple learn, guess, slip, or forget rates, a subscript $P(T_i)$, $P(G_i)$, $P(S_i)$, $P(F_i)$ denotes the corresponding probability, or rate, for class $i = 1 ... m$, respectively.

Initial and fit BKT model parameters are represented using a Python dictionary. Inside of this dictionary, we store $A$, which is a collection of matrices with each 2x2 matrix corresponding to the learning and forgetting probabilities for each learn class in order to aid in efficient matrix multiplication during fitting. $A$ has the format where $m$ is the total number of learn rates: \[
\begin{bmatrix}
\begin{bmatrix}
    P(\neg{T}_1)  &  P(F_1)      \\
    P(T_1)  &  P(\neg{F}_1)      
\end{bmatrix}
, \cdots,
\begin{bmatrix}
    P(\neg{T}_m)  &  P(F_m)      \\
    P(T_m)  &  P(\neg{F}_m)       
\end{bmatrix} 
\end{bmatrix}
\]
We define $\alpha$ as a set of 2-length vectors each corresponding to $[P(\neg{L}_t), P(L_t)]$ for all time steps for a specific student. Similarly, $\pi_0$ stores information about the prior, in the format of $P(\neg{L}_0), P(L_0)$. 

\subsubsection{EM and BKT Algorithm}
We use Expectation Maximization to fit model parameters, shown to provide desirable convergence properties, given plausible initial parameter values \cite{pardos2010navigating}. Inside the EM and inference algorithms, we use several intermediate data structures and vectorization to improve computational efficiency in fitting the models. To calculate $\alpha[t+1]$ given $\alpha[t]$, we multiply it by the part of the learn/forget transition matrix $A$ corresponding to the learn class of time step $t$. We element-wise multiply by the vector likelihoods which consists of $[P(G), P(\neg{S})]$ or $[P(\neg{G}), P(S)]$ for the corresponding guess class of time step $t$, depending on whether the student answers correctly or not, respectively. Finally, normalizing this vector results in $\alpha[t+1]$. We demonstrate the algorithm for an example iteration of the BKT algorithm with learn class 1, guess class 1, and an incorrect response observed ($obs = 0$) at time step $t$.

$$
{
\begin{bmatrix}
    P(\neg{L}_{t+1})\\
    P(L_{t+1})
\end{bmatrix} 
}
=
\begin{bmatrix}
    P(\neg{T}_1)  &  P(F_1)\\
    P(T_1)  &  P(\neg{F}_1)
\end{bmatrix}(
\begin{bmatrix}
    P(\neg{L}_t)\\
    P(L_t)
\end{bmatrix} 
\circ
\begin{bmatrix}
    P(\neg{G}_1)\\
    P(S_1)
\end{bmatrix})
$$
$$
=
\begin{bmatrix}
    P(\neg{T}_1)*P(\neg{L}_t|obs_t) + P(F_1)*P(L_t|obs_t)      \\
    P(T_1)*P(\neg{L}_t|obs_t) + P(\neg{F}_1)*P(L_t|obs_t)   
\end{bmatrix} 
$$

At the end of this $\alpha$ calculation, we perform the E-step of the EM algorithm by recursively calculating an expectation $\gamma$ for each time step by backtracking through the learned latent states. We can then take the global average of the expectations of the learn/forget transition matrices, guess/slip vectors, and priors during the M-step and use these as the parameters for the next iteration of EM. In terms of the number of students $S$ and the typical sequence length for each student $T$, the model fitting algorithm's asymptotic time complexity for standard BKT is $\Theta(TS)$.

\subsubsection{C++/Python Implementation Details}
We use a C++ extension to perform the EM iterative updates and matrix multiplication for the model fit and prediction process. This allows us to use efficient linear algebra libraries in C++\footnote{Boost was previously used as a connector between Python and the C++ extension, but it has been deprecated since pyBKT 1.2.2, resulting in a 3-5x performance increase.} and benefit from greater support for multithreading through OpenMP. 

We use Eigen to perform the matrix operations. There are many technical advantages of using Eigen with a linear algebra heavy model such as pyBKT. Eigen provides efficient, thread-safe matrices and arrays, while being a relatively portable package distributed along with pyBKT. It allows for lazy evaluation, expression templates and compiler optimizations. 

We use OpenMP for parallelizing the demanding model fitting process. OpenMP is a universally accepted multithreading library for C/C++ that exploits multicore processors with low overhead. With a shared memory space for forked threads, OpenMP avoids overhead for inter-process communication (IPC) unlike Python's multiprocessing. With Eigen's explicit support for OpenMP-based multithreading, heavy matrix operations and iterative processes are further optimized in pyBKT.

\subsubsection{Pure Python Implementation Details}
We wish to maintain the accessibility of pyBKT across all platforms while maintaining as much efficiency as possible. To do this, a pure Python implementation, without C++ extensions, is included. This implementation provides quick access to any user, including on Windows, wishing to run a BKT model without the hassles of compilation and complex dependencies. We relax this version of the library's requirements to include mostly native modules along with the widely supported NumPy.

NumPy is used for the matrix operations in the pure Python build of pyBKT as it is the most efficient and widely used numeric computational library in Python. Technically, it provides impressive single-threaded performance. Similarly to other optimized mathematical libraries such as Eigen, NumPy employs code vectorization, efficient memory mapping techniques for sparse matrices, and compiler optimizations.

Since NumPy is primarily a single-threaded application with little support for multicore scaling, we use Python's native multiprocessing library for parallelizing the model fitting. It provides native CPython support for multicore scaling to bypass the Global Interpreter Lock (GIL), which allows only one running thread of execution within a process. Although different Python implementations (i.e JPython) exist to disable the GIL or remove the memory overhead, we use a native module for simplicity and speed. 

\subsection{Runtime Evaluation}

\begin{table*}[h]
\centering
\caption{Comparison of runtimes, scale factor and speedup between Python and C++/Python implementations of pyBKT.}
\begin{tabular}{|l|l|l|l|l|l|}
\hline
\textbf{Test Description}                                 & \multicolumn{2}{c|}{\textbf{Pure Python}} & \multicolumn{2}{c|}{\textbf{C++/Python}} & \textbf{Speedup} \\ \cline{2-5}
\textbf{}                                                 & Runtime       & Scaling                   & Runtime     & Scaling                    & \textbf{}        \\ \hline
Synthetic Data Generation, Model Fit (500 students)       & 160.12s      & \multicolumn{1}{c|}{}     & 1.07s       & \multicolumn{1}{c|}{}      & 149.64x          \\ \cline{1-2} \cline{4-4} \cline{6-6} 
Synthetic Data Generation, Model Fit (5,000 students)     & 1,596.30s     & 9.97x                     & 2.62s       & 2.45x                      & 609.27x          \\ \hline
Synthetic Data Generation, Model Predict (500 students)   & 8.02s         & \multicolumn{1}{c|}{}     & 0.50s       & \multicolumn{1}{c|}{}      & 16.04x           \\ \cline{1-2} \cline{4-4} \cline{6-6} 
Synthetic Data Generation, Model Predict (5,000 students) & 67.08s       & 8.36x                     & 2.42s       & 4.84x                      & 27.71x           \\ \hline
Cross-validation, Cognitive Tutor                         & 320.28s      & \multicolumn{1}{c|}{-}    & 4.79s       & \multicolumn{1}{c|}{-}     & 66.86x           \\ \hline
\end{tabular}
\end{table*}


We compare the runtime performance of the pure Python and C++/Python implementations of pyBKT on five typical model fitting and prediction tasks. We present two sets of tasks, fitting and prediction on synthetic data, that additionally showcases the way in which the runtimes scale with the size of input data for both implementations of pyBKT. Each of the tasks are averaged over several runs for both implementations of pyBKT on a machine with 2 x Intel(R) Xeon(R) CPU E5-2620 v3 CPUs at 2.4Ghz with 256GB of system RAM. The results are shown in Table 1.

We evaluate the runtimes using two metrics. The scaling factor is defined as the ratio of the runtime of the larger input and the smaller input for a set of prediction or fitting tasks. The speedup is defined as the ratio of the runtime of our C++/Python implementation to our pure Python implementation.

The first four tasks perform synthetic data generation for 500 students and 5,000 students respectively with a sequence length fixed at 100 followed by prediction or fitting. These tasks illustrate a typical medium and large workload for model fitting and prediction tasks. The generated synthetic data is fit or predicted using a standard BKT model. It is clear that as the number of students scales, the pure Python implementation of pyBKT performs and scales more poorly with the number of students. The C++/Python implementation shows a nearly 150-600x speedup for fitting and 15-30x speedup for prediction as compared to the pure Python version. In comparison to its predecessor xBKT (MATLAB), the C++/Python version of pyBKT gains a 3-4x speedup across all fitting and prediction tasks. In \citet{xu2015xbkt}, it is noted that xBKT outperforms BNT by 10,000x, which suggests a 30,000-40,000x speedup of pyBKT as compared to BNT.

The final task performs a cross-validated prediction task for a selected skill in the Cognitive Tutor dataset. We use a variety of models (standard, multigs, multiprior) to test predictive accuracy and measure its runtime. This task is around 65x slower in the pure Python implementation.

While the runtimes and the overall scaling of the pure Python port with respect to the size of input data are significantly poorer for each task, that is an expected trade-off with regards to accessibility and portability. For an end-user that is training and testing moderately-sized BKT models or evaluating models, they would benefit from a portable and universal BKT model which can handle a moderate input size with relative efficiency. For heavier research-oriented or production-oriented tasks, the C++/Python implementation is recommended since it generally performs much more efficiently.

\section{Data Sufficiency Analysis}
We examine the data sufficiency requirements of the standard BKT model by exploring trade-offs between input size and parameter error and mastery estimation accuracy. We define the input size as the magnitude of the number of students and the average sequence length. Through the first analysis, practitioners may gain an intuition for the minimum cohort size and minimum number of questions answered per student per skill to effectively apply BKT. Our second analysis in this section focuses on mastery estimation accuracy, also using synthetic data. This analysis depicts how the worst-case expected mastery estimation accuracy decreases as a function of sequence length for a given set of prior, guess, and slip parameters. 


For the following analyses, we generate synthetic data, both responses and mastery states, from pyBKT using ground truth parameter values set to common values seen for Algebra skills\footnote{prior=0.08, guess=0.15, slip=0.05, forget=0 and learn=0.3 for the first analysis and variable for the second}. In doing so, we are able to calculate the error of the fit parameters and accuracy of the mastery estimation.  

\subsection{Synthetic Model Fit Accuracy}

The synthetic generation of data is performed for input sizes from 10 to 200 students for all sequence lengths from 2 to 35. While 200 is not an uncommon number of students to have in a cohort, more than 30 responses to a single skill would be unusual for a student. Each combination of input size is averaged over five fits, each of which includes the best model over 20 random EM fit initializations. 

The error of the model's fit parameters are analyzed using the mean absolute percentage error (MAPE) in relation to predefined ground truth generating parameters. We plot the fitting error of the learn, slip, guess, and prior parameters as a function of the number of synthesized students (Figure 1, left) and length of synthesized response sequences for each student (Figure 1, right). While all data points cannot be visualized on a single plot, we show the data points for the prior parameter as an example.

For all parameters, there is a clear negative and exponential error decay with respect to the number of students. This is consistent with an expected asymptotic behaviour when increasing the number of students in the fitting procedure. Learn and slip parameters asymptote at around 50 students while guess and prior do so after 100, given a sequence length of 10. 


There seems to be a slowly decreasing linear relationship between the typical sequence length and parameter fit error, with the prior parameter showing the greatest improvement in MAPE. These analyses show that there is not nearly as much benefit to fitting error by increasing the sequence length (i.e., giving students more problems) as there is by increasing the number of students.

\begin{figure}[h!]
\caption{Mean absolute percentage error (MAPE) of fit parameters as a function of number of students (left) and sequence length (right).}
\centering
\includegraphics[width=8.5cm]{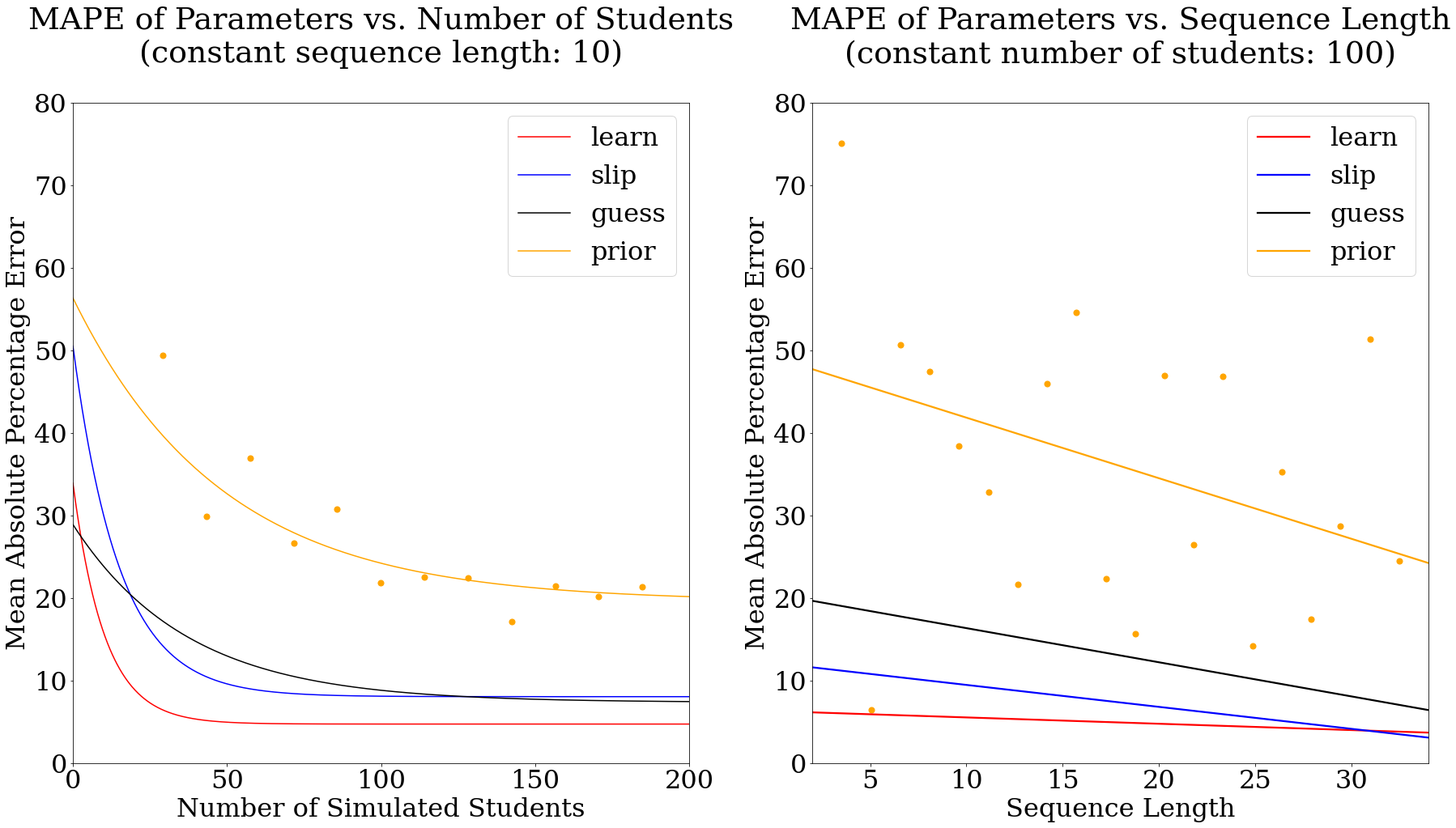}

\end{figure}

\subsection{Synthetic Mastery Estimation Accuracy}
We look to evaluate the worst-case accuracy of the standard BKT model's mastery estimation, its most common task within a computer tutoring system, using a mastery threshold of $P(L_{t}) \geq 0.95$ on simulated problem solving sequences. We use the same predetermined ground truth BKT parameters and data generation methodology as in the previous subsection with the exception of the learn rate, which is set dynamically in this analysis. 

It is known that the probability of mastery will converge to $1 - forgets$ in a finite number of time steps given a learn rate > 0 \cite{van2013properties}. This means that cognitive mastery estimation accuracy will increase with respect to sequence length, as student mastery state becomes more homogeneous.

In order to model the worst-case mastery estimation of the model, we find the learn rate that will produce an average probability of mastery close to 0.5 across students at the final time step $\tau$ for all our chosen sequence lengths, thus preventing a trivial estimation of a majority mastery state. We find the learn rate via grid search with a granularity of 0.001 since this cannot be solved analytically \cite{van2013properties}.


The mastery estimation accuracy exponentially decays upward with respect to sequence length with a Pearson correlation R\textsubscript{log(acc)} = 0.73 as shown in Figure 2. The mastery estimation accuracy in our analysis can be observed to asymptote around a sequence length of 15. This suggests that the worst-case mastery estimation accuracy scenario can be mitigated, given our chosen predefined parameter values, with an average response sequence length per skill of 15 or greater.

\begin{figure}[h!]
\caption{Worst-case accuracy of mastery estimation as a function of sequence length for the standard BKT model.}
\centering
\includegraphics[width=8cm]{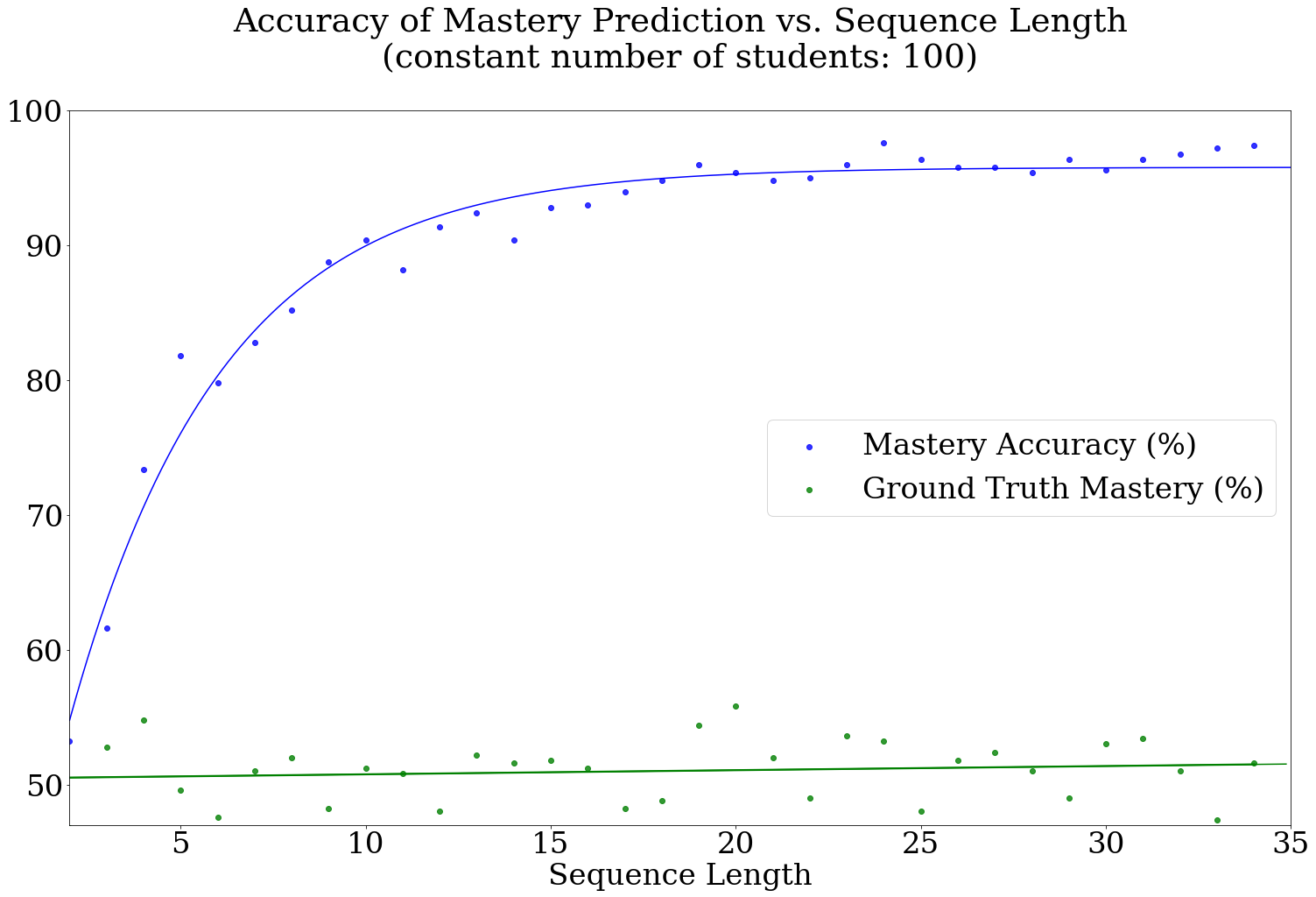}
\end{figure}

\section{EVALUATION OF PYBKT MODELS}\label{sec:Evaluation}

To gauge the validity of the BKT models and extensions supported by pyBKT, we perform predictive model evaluations replicating those found in prior literature. Models were evaluated by performing five-fold cross-validation on the ASSISTments 2009-2010 Skill Builder dataset, the Cognitive Tutor 2006-2007 Bridge to Algebra dataset\footnote{\url{https://pslcdatashop.web.cmu.edu/KDDCup}}, and specialized datasets from \citet{feng2008can} and \citet{piech2015deep} in order to validate more specific model extensions. While there are slight differences between our results and the results of other papers, we believe this is due to our random parameter initialization and small differences in fitting methods. Our complete model evaluations, data, and results are located in a pyBKT examples repository\footnote{\url{https://github.com/CAHLR/pyBKT-examples}}.

\subsection{Standard BKT and BKT+Forget}
In general, the predictive accuracy of the models generated by pyBKT are in line with what others have observed from BKT. For instance, in \citet{khajah2016deep}, they found that by fitting the basic BKT model on the train/test split provided by \citet{piech2015deep} from the 2009-2010 ASSISTments dataset for each skill, they obtained an AUC of 0.73, whereas pyBKT predicts with an AUC of 0.76. Similarly, when \citet{khajah2016deep} added the forgets parameter into their model, they achieved an AUC of 0.83, exactly the same AUC we achieve using pyBKT.

\subsection{KT-IDEM}
When using the KT-IDEM model on the ASSISTments 2009-2010 Skill Builder dataset, pyBKT achieves an average AUC increase of 0.01932, which is very close to the 0.021 average increase reported in \citet{pardos2011kt}. While the exact subset of data was not exactly specified in that paper, we used the ten skills with the most responses, using different template ids as the guess/slip classes as prescribed in \cite{pardos2011kt}. Since the ASSISTments data has a very high average response to template ratio ($\sim$1,000), the KT-IDEM model performs very well compared to the standard BKT model using RMSE as the metric of comparison, being lower or equal in nine of the ten skills selected. 

\subsection{KT-PPS}
The Prior Per Student model was applied to the ASSISTments' Groups of Learning Opportunities dataset \cite{feng2008can} consisting of 42 problem sets. 
In \citeauthor{pardos2010modeling}, it was found that the KT-PPS model performed better than the standard BKT model on 30 out of the 42 problems sets, as evaluated on the predictions of the last response of each student's response sequence. This was achieved using a variant of KT-PPS that models a high and low prior and assigns students to the high prior if they answer correctly on the first problem of the problem set, and to the low prior otherwise. The high prior was set to an ad-hoc value of 0.90 and the low prior to 0.15 in that work. 

The pyBKT replication of this model is done without true multiple prior modeling. Instead, when the multiprior option is set to True, $P(L_0)$ is set to 0 and a dummy time step is created at the beginning of the sequence. Three learn rates are created, the first corresponding to the high prior, the second to the low prior, and the third corresponding to the standard $P(T)$ applied between all subsequent time steps. The initial values of these virtual priors were set to the ad-hoc values from \cite{pardos2010modeling}; however, since pyBKT does not support parameter fixing as of this writing, these parameters were learnable. With these settings, pyBKT's KT-PPS performs better than standard BKT on 27 out of 42 of the problem sets. The small difference in prediction accuracy of this model may be attributable to the difference in the algorithm regarding fixed parameters, but the similarity in performance is promising.

\subsection{Item Order and Item Learning Effect}
Results from the Item Order Effect \cite{pardos2009determining} and Item Learning Effect \cite{pardos2009detecting} papers were not focused on response prediction improvement. In fact, no prediction accuracy results were provided. Instead, the purpose of the models was to compare the learn rates of classes to flag effective and ineffective items and orders. The examples repo of pyBKT depicts such differences. Nevertheless, a modest 0.01 RMSE improvement for both model variants was obtained compared to the standard BKT model. 

\section{Conclusions}
We introduced pyBKT as a seamlessly installable, efficient, and portable Python library with model extensions such as KT-IDEM, KT-PPS, BKT+Forgets, Item Order Effect and Item Learning Effect. The Model class abstraction in pyBKT provides an expressive way to interact with the BKT model extensions with ease, with one-line methods to create, initialize, fit, predict, evaluate, and cross-validate any combination of BKT model extensions. We measured the runtime of pyBKT to be nearly 3x-4x faster than its predecessor, xBKT, and nearly 30,000x faster than BNT, a standard BKT implementation. Through the analyses presented, we established 50 as a reasonable number of students to achieve convergence to canonical parameter values with any average student sequence length and 15 as a reasonable sequence length to mitigate worst-case mastery estimation accuracy. Lastly, through real-world dataset analyses, we showed the validity of the model implementation through its agreement with past results using established software. 


\section*{Acknowledgments}
We recognize Matthew J. Johnson, who co-developed xBKT, the predecessor to pyBKT, and Cristian Garay, who developed the initial Python and Boost adaptation of xBKT.

%
{\footnotesize\bibliography{sigproc}}  
\end{document}